\documentclass[runningheads,a4paper]{llncs}
\usepackage[bookmarks=false,hidelinks]{hyperref}
\usepackage[T1]{fontenc} 
\usepackage[utf8]{inputenc} 
\usepackage{amssymb}
\usepackage{amsmath}
\usepackage{graphicx}
\usepackage{url}
\usepackage{color}
\usepackage{apacite}
\usepackage{appendix}
\newcommand{\keywords}[1]{\par\addvspace\baselineskip
\noindent\keywordname\enspace\ignorespaces#1}

\newcommand{\vq}{\textsc{VQ}}
\newcommand{\vae}{\textsc{VAE}}
\newcommand{\vqvae}{\textsc{VQ-VAE}}
\newcommand{\vqcpc}{\textsc{VQ-CPC}}
\newcommand{\vqvaetwo}{\textsc{VQ-VAE-2}}
\newcommand{\gansynth}{GANSynth}
\newcommand{\wavenet}{WaveNet}

\begin{document}

\mainmatter  

\title{Spectrogram Inpainting for Interactive Generation of Instrument Sounds}


%
%
\author{Th\'eis Bazin\inst{1,2}\thanks{This author was supported by the ANRT via the CIFRE contract number 2019.0094.}\and{}Ga\"etan Hadjeres\inst{1}\and{}Philippe Esling\inst{2} \and{}Mikhail Malt\inst{2}}
%
\authorrunning{Th\'eis Bazin \and Ga\"etan Hadjeres \and Philippe Esling \and Mikhail Malt}

\institute{Sony CSL Paris \and IRCAM - CNRS UMR 9912 STMS - Sorbonne Universit\'e}

%
%

\maketitle
\begin{abstract}

Modern approaches to sound synthesis using deep neural networks are hard to control, especially when fine-grained conditioning information is not available, hindering their adoption by musicians.

In this paper, we cast the generation of individual instrumental notes as an inpainting-based task, introducing novel and unique ways to iteratively shape sounds. To this end, we propose a two-step approach: first, we adapt the \vqvaetwo{} image generation architecture to spectrograms in order to convert real-valued spectrograms into compact discrete codemaps, we then implement token-masked Transformers for the inpainting-based generation of these codemaps.

We apply the proposed architecture on the NSynth dataset on masked resampling tasks. Most crucially, we open-source an interactive web interface to transform sounds by inpainting, for artists and practitioners alike, opening up to new, creative uses.


\end{abstract}
\section{Introduction\label{sec:introduction}}

Generative modeling of audio has seen a surge in fidelity in recent years, thanks to advances in generative neural network-based architectures~\cite{van_den_oord16:arxiv:wavenet,van_den_oord16:pmlr:parallel_wavenet,engel19:iclr:gansynth,aouameur:hal-02479211,DBLP:journals/corr/abs-1907-00971}. These new models are progressively bridging the gap with dedicated synthesis software in terms of the versatility of sounds produced without requiring the involved domain-specific knowledge of audio synthesis. Yet these approaches still fall short of providing convenient, interpretable control over the generated sounds, only giving access either to \emph{global} parameters that change the texture of the whole sounds or to \emph{dense}, high-frequency conditioning, impractical for interactive uses. 





\emph{Global} conditioning labels such as pitch and instrument family set aside, 
introducing \emph{local} control in such models is made difficult by the typically high sampling frequency of sounds in waveform representation. Indeed, this high sampling frequency makes autoregressive modeling of sounds highly challenging. This is the case for models of the \wavenet{} family~\cite{van_den_oord16:arxiv:wavenet,van_den_oord16:pmlr:parallel_wavenet}, and such models are bound to operate with low receptive fields and display highly unstationary behaviour. Controlling these models therefore typically requires very dense conditioning sequences, sampled at a rate close to the audio rate, which is impractical for end-users. Conversely, models such as \gansynth{}~\cite{engel19:iclr:gansynth}, that use spectrogram representations and generate full-scale sounds of several seconds in parallel through transposed convolutions conditioned on a single vector, lose the ability to control the small-scale structure of the sounds.

Promising solutions have been proposed to introduce control at a mid-level range in autoregressive models through the use of Style Tokens and similar techniques~\cite{engel17:icml:neural_audio_synthesis_of_musical_notes_with_wavenet_autoencoders, wang18:iclr:style_tokens}. Here, an intermediary, subsampled latent space is learned along with the autoregressive decoder and used to condition the decoder. Nevertheless, the resulting conditioning vectors are real-valued vectors with unindentified dimensions, making them hard to interpret. Existing control schemes are therefore limited to extracting the sequence of conditioning vectors of an example sound and transferring it to another sound, but further control is difficult.

We turn to recent advances in \emph{inpainting}-based interactive generation, which have brought new approaches to controllable sampling for creative applications. Inpainting models are prediction models trained to reconstruct hidden or missing data in an input.
These models were previously proposed for the interactive generation of symbolic music~\cite{hadjeres17:icml:deepbach,hadjeres18:anticipation_rnn} and are furthermore readily amenable to interactive applications~\cite{bazin19:iccc:nonoto}.

In this work, we propose to tackle the problem of controlled sound generation through a two-steps approach. First, an encoder-decoder network is trained to compute a highly compressed representation of the sounds through downsampling and discretization, with efficient encoding as well as decoding. This mid-level representation, used as input for our subsequent generative models, bridges the gap between the dense conditioning sequences of \wavenet{} and the very sparse, label-based conditioning of \gansynth{}. We then train powerful Transformer-based sequence generation models on these discrete codemaps for interactive generation by inpainting.

\section{Contributions\label{subsec:contributions}}

We adapt the \vqvaetwo{} image generation model to allow for the interactive generation of sounds via inpainting. Our key contributions are:
\begin{enumerate}
    \item We cast the generation of musical instrument sounds as an interactive, inpainting-based generative-modeling task,
    \item We adapt the hierarchical \vqvaetwo{} model to operate on spectrograms, using a custom modification of the IF-Mel spectrogram-based representation originally proposed for the \gansynth{} model,
    \item In order to sample new sounds in this compact latent space, we model a factorization of the joint probability of hierarchical codemaps from the \vqvaetwo{}. To this end we introduce hierarchically conditioned sequence-masked Spectrogram Transformers, compatible with interactive generation by inpainting. These models furthermore support conditioning through global labels for additional control,
    \item We open-source the code for this model and release an open-source interactive web interface for the proposed model\footnote{Code for the models and interface available at: \url{https://github.com/SonyCSLParis/interactive-spectrogram-inpainting}}, which allows to edit sounds via inpainting whilst viewing and listening to the results, enabling novel and unique schemes of sound transformation.
\end{enumerate}

\section{Proposed approach}

In order to provide control on the generated sounds, a sweet spot must be found between full autoregressive sampling of full-scale spectrograms, which is computationally demanding when using powerful Transformer architectures, and one-shot parallel generation \emph{à-la}-\gansynth{}, which discards local control in favor of computational efficiency.
We therefore follow an autoregressive approach but introduce models with highly downsampled and compact latent representations, so that these latent sequences can in turn be conveniently generated and edited using autoregressive models.

\subsection{Compact representations of sounds}

\begin{figure}[tpbh]
\centering
  \includegraphics[width=0.99\columnwidth]{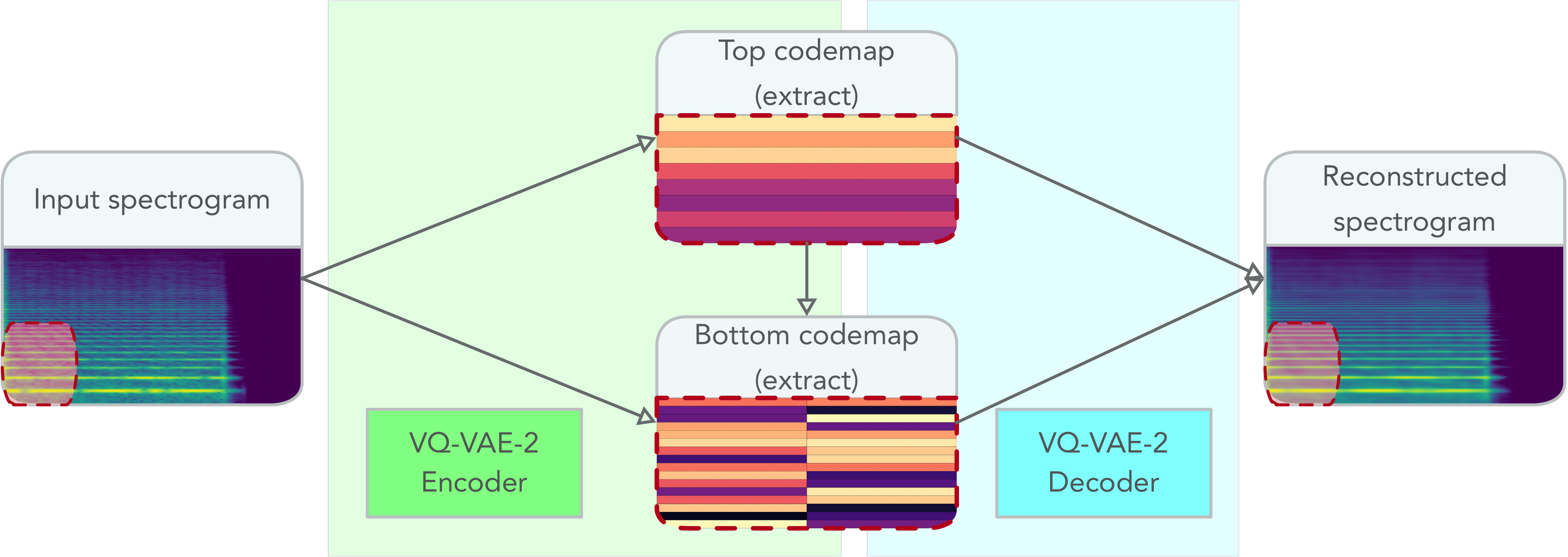}
  \caption{Proposed VQ-VAE-2 architecture for spectrograms.
  The encoder converts  \( 1024 \times 128 \)-large spectrograms into two compact integer maps, \( c_{top} \) and \( c_{bottom} \) of size \( 32 \times 4 \) and \( 64 \times 8 \). The decoder is trained to reconstruct the original spectrogram from these discrete codemaps.\label{fig:vqvae}}
\end{figure}

The \vqvaetwo{} model, initially proposed for the generation of images~\cite{razavi19:neurips:generating_diverse_high_fidelity_images_with_vqvae_2}, defines an autoencoding architecture using a hierarchy of discrete latent variables. In the \vqvaetwo{} model, the \vae{}'s latent space is made discrete through \emph{vector-quantization}, a classic dictionary-learning algorithm~\cite{gray:ieee1994:vector_quantization}: the latent representation computed by the \vqvaetwo{} encoder is projected onto a learned \emph{codebook} (a dictionary of code vectors), effectively replacing the input data by a discrete grid of codebook indexes. The decoder in turn receives only the projection and reconstructs the original input by upsampling (via transposed convolutions).
In the hierarchical approach, a top codemap is obtained from the bottom map by further downsampling and vector-quantization.  In this setting, the highly downsampled top codes can be interpreted as depicting the overall, large-scale structure of the images. The bottom codemap decoder is subsequently conditioned on this global structure and refines it with fine-grained detail. This is shown on \autoref{fig:vqvae} via the arrow depicting the conditional dependency of the bottom codemap on the top one.


We note that \vq{} has previously been applied to the generation of audio signals with impressive results~\cite{dieleman18:nips:challenge_raw_audio_at_scale,vandenoord17:nips:neural_discrete_representation_learning_aka_vqvae}, most notably in the recent Jukebox model by Open AI~\cite{dhariwal2020:arxiv:openai_jukebox}. Nonetheless, \vq{} is used in these approaches as a powerful information bottleneck allowing to scale autoregressive models of audio to very large temporal scales of several tens of seconds. Controlling the huge models proposed in these papers remains highly challenging, if only due to the very high computational load they incur, making these approaches somewhat orthogonal to our goal of intuitive and interactive control.



\subsection{Spectrogram representations for intuitive control}

Since the \vqvaetwo{} model was initially introduced for the generation of images, we borrow the invertible Hi-Res Mel-IF (Mel-Amplitude and Instantaneous Frequency) spectrogram representation that was shown by~\citeA{engel19:iclr:gansynth} to provide the best results for generation with convolutional neural networks. This representation, which performs a temporal unrolling of the phase channel of the spectrograms, introduces smoothness and stability as opposed to directly learning to model the phase. This representation can furthermore be inverted back to the original spectrogram at no cost and therefore allows for efficient conversion back to audio using the inverse Fourier transform.

The auto-encoding training of the \vqvaetwo{} furthermore requires introducing a reconstruction loss, as opposed to the adversarial training of \gansynth{}. We propose to use a slightly modified \(L2\)-loss, using a phase thresholding process: we ignore IF reconstruction errors (by setting those gradients to zero) in areas of low sound amplitude, where the IF starts becoming highly noisy~\cite{engel19:iclr:gansynth}, effectively shifting gradient signals to areas with significant amplitude.

Working on spectrograms has the additional advantage of separating time and frequencies onto two separate axis in the input representation. This provides an intuitive interpretation of the latent codes learned by the hierarchical \vqvaetwo{} as time-frequency atoms. Editing the resulting 2D codemaps by inpainting can then be seen as performing localized transformations in the time-frequency plane over multiple frequency bands, akin to the multi-band equalizers musicians are familiar with.


\begin{figure*}[tpbh]
  \centering
  \includegraphics[width=0.99\textwidth]{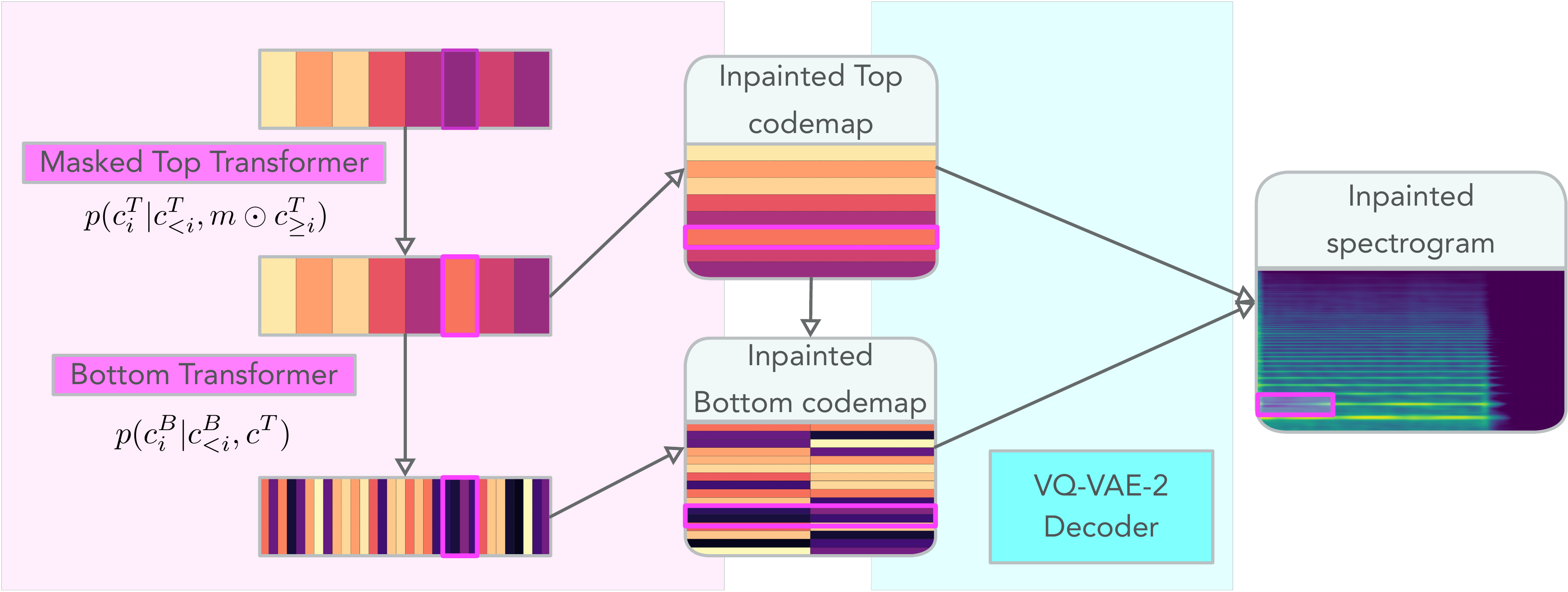}
  \caption{Diagram of the proposed spectrogram inpainting process.
  After a linearization-procedure informed by harmonic structure, sequence-masked Transformers are trained to perform inpainting on the codemaps of the \vqvaetwo{}, which allows to sample new sounds by first autoregressively sampling from the factorized distribution \( p (c_{top}) p(c_{bottom} | c_{top}) \) then decoding these sequences.
  \label{fig:transformers}}
\end{figure*}

\subsection{Spectrogram Transformers}

After training the \vqvae{}, the continuous-valued spectrograms can be replaced with the hierarchical latent codemaps. Generating new sounds then amounts to modeling and sampling from the joint probability \( p(c_{top}, c_{bottom})\). We follow the approach of \cite{razavi19:neurips:generating_diverse_high_fidelity_images_with_vqvae_2} and factorize this probability into \( p(c_{top}) p(c_{bottom}|c_{top})\), which is also in line with the conditional structure of the \vqvaetwo{} encoder. We therefore train an autoregressive model for the top codes and another one -- conditioned on the top codes -- for the bottom codes.

To this end we adapt the \vqcpc{} architecture~\cite{hadjeres20:vqcpc} for hierarchical discrete latent variables. For the top codemap \(c^{T}_{i\dots{}N}\), we use a causally masked, self-attentive Transformer, therefore modeling the probability \(p(c^{T}_i|c^{T}_{<i}, m \odot{} c^{T}_{\geq{}i})\), where \( m \) denotes the randomly sampled \emph{inpainting mask}, used to stochastically hide data during training. Typically, a simple Bernoulli can be used here for each token. This mask then allows, at inference time, to select areas to regenerate, and provide the model with information about the fixed, future tokens, ensuring coherent generation.
For the upsampled, bottom codemaps we use another Transformer which can attend via attention to the whole top codemap in a patch-based fashion: indeed, to help the attention procedure, we align patches of the bottom map to the top code which they were upsampled from. We obtain the following formulation: \( p(c^{B}|c^{T}) = \prod_{i}p(c^{B}_i|c^{B}_{<i}, c^{T}) \).
The resulting architecture is depicted on \autoref{fig:transformers}.

\section{Experiments}

The proposed hierarchical model allows to perform various generative modeling tasks, we provide some usage examples here. We refer the reader to the companion website\footnote{\url{https://sonycslparis.github.io/interactive-spectrogram-inpainting/}} where many uncurated audio samples are provided for each task described here, using all of the instrument types in the NSynth dataset and across a large range of pitches, along with a more thorough discussion of the performances of the model. Note that in this paper, we focus on the general framework of interactive sound inpainting and its applications and we reserve a quantitative evaluation of architectural choices for future work.

\subsection{Training}

We train the proposed architecture on the NSynth dataset~\cite{engel17:icml:neural_audio_synthesis_of_musical_notes_with_wavenet_autoencoders}. As is done in the \gansynth{} paper~\cite{engel19:iclr:gansynth}, we restrict the dataset to sounds with pitch from MIDI 24-84, avoiding extreme pitches. 
We compute the input spectrograms using windows of 2048 samples with \( N_{fft} = 2048\) with a \( 75\% \) overlap factor, resulting in an effective dimension of \( 1024 \times 128 \) for the spectrograms.
As opposed to the representation used in \gansynth{}, we lower the Mel-scale break frequency from \( 700 \)Hz to \( 240 \)Hz, in order to increase resolution in the low frequencies, bringing more precision in the interactive editing of the lower harmonics.

The \vqvae{} is trained for reconstruction of the Mel-IF representation using the phase-masked L2 loss described above with the ADAM optimizer. After training the \vqvae{}, we use its encoder to extract the latent codemaps for all of the sounds in the training dataset. The bottom and top Transformers can then be independently trained on these codemaps. Here, we train in an autoregressive fashion with the RAdam optimizer~\cite{liu2019:iclr:radam} and a label-smoothed prediction loss~\cite{muller19:neurips:when_does_label_smoothing_help}. 
More details are given in the repository accompanying the paper.

\subsection{Results}

\begin{figure}[hpbt]
 \centering
  \includegraphics[width=0.8\columnwidth]{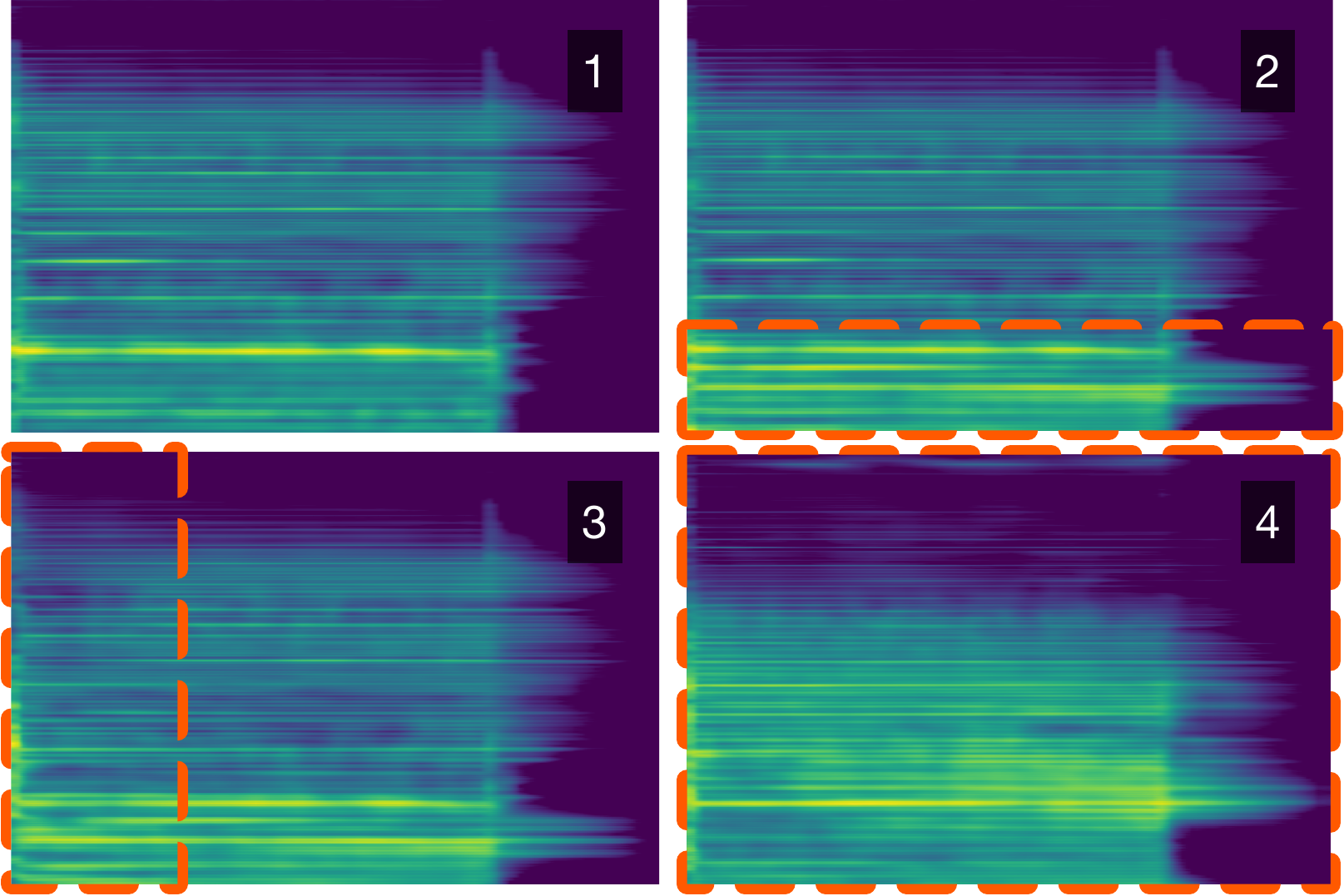}
  \caption{Successive inpainting operations over a generated sample. (1) Generated sample ; (2) inpainting of the lower frequencies over the whole duration of the sound ; (3) inpainting of the full first second ; (4) inpainting of the full bottom codemap with the top codemap fixed.\label{fig:inpainting}}
\end{figure}

\subsubsection{Inpainting operations\label{subsec:inpainting_operations}}

On~\autoref{fig:inpainting} we display uncurated samples for three possible uses of the model. First, a spectrogram for a 4 second organ sound at pitch 60 generated by our model is shown. Three successive inpainting operations are then presented. The first two examples operate on both codemaps, by first autoregressively resampling the displayed portion of the top codemap, then resampling the aligned tokens in the bottom codemap. The third one only regenerates the bottom codemap, keeping the top one fixed. In the first example, the five lowest frequency bands of the top map are regenerated over the whole duration of the sound with constraint \( \texttt{"Bass + pitch C1"} \), resulting in a visible increase in lower-frequency energy. Then, the first second of the spectrogram is regenerated over all frequency bands with constraint \( \texttt{"Mallet + pitch C3"} \), resulting in a more marked attack. In the last step, the top codemap is kept fixed but the full bottom codemap is resampled conditioned on it, with a constraint  \( \texttt{"Synth Lead + pitch C5"} \). We can see that the global structure of the spectrogram remains similar, but the local timbral texture indeed changes. We again refer the reader to the accompanying webpage for audio examples.



\subsubsection{Interactive interface}

\begin{figure}[hpbt]
\centering
  \includegraphics[width=0.75\columnwidth]{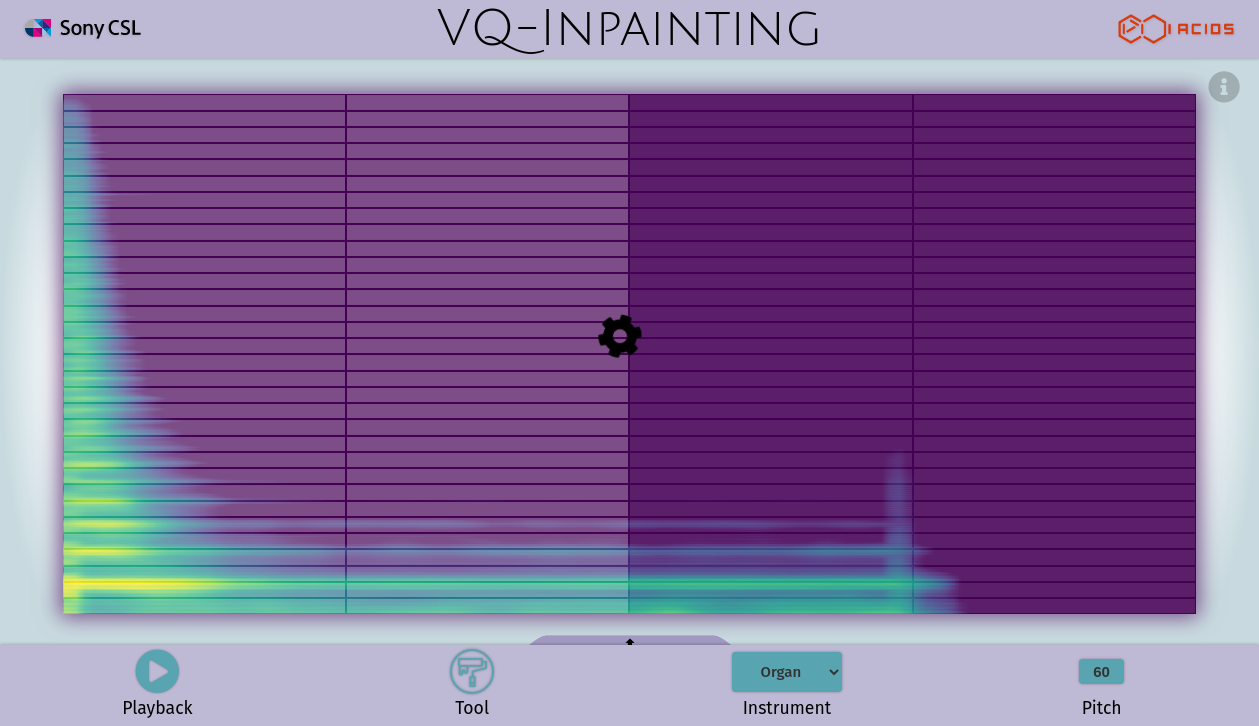}
  \caption{Screenshot of the proposed interactive web interface. The highlighted zone is currently being regenerated by the back-end.\label{fig:interface}}
\end{figure}

We believe that generative models for music only reveal their potential through actual usage in music production contexts, which requires providing musicians with intuitive interfaces to these complex models. In line with this philosophy, we propose a TypeScript-based web interface that allows to generate and edit sounds using the models presented in this paper\footnote{Available within the code's repository at \url{https://github.com/SonyCSLParis/interactive-spectrogram-inpainting}}, using a PyTorch-based computing back-end. This interface is represented on \autoref{fig:interface}.

The \emph{modus operandi} is a simple one, akin to classic image editing software, which we can expect our users to be familiar with. An initial sound is either generated from scratch using the autoregressive models, sampled from a dataset or provided by the user through drag and drop. The \vqvae{}-reconstructed spectrogram is displayed on the interface, along with a grid overlay depicting the cells of the \vqvae{}'s latent codemaps. The user can then perform inpainting operations as described in \ref{subsec:inpainting_operations} by selecting zones on this grid, either on the coarse top codemap or on the more detailed bottom map.

We note that the possibility of using different instrument and pitch constraints for successive inpainting operations opens the way to the generation of heterogeneous spectrograms, unseen in the training dataset -- such as inpainting the attack of an organ sound with a \texttt{"Guitar"} instrument constraint to obtain a plucked organ. The successive inpainting operations performed in~\autoref{fig:inpainting} are an example of this use case.
We finally observe that, due to the inherently linearly-scaling nature of the autoregressive generation, each token-wise prediction of the spectrogram transformers should be fast if responsive interaction is desired. Currently, typical local operations (over a region of one second) take around one second to complete, and work is undergoing to reduce this latency for a seamless interactive process.

\section{Conclusion and perspectives}

We have presented an approach to interactively generating instrument sounds by inpainting. To this end, we adapted the \vqvaetwo{} image generation model to spectrograms via a representation inspired by the Mel-IF representation from \gansynth{}. This allows to encode sounds into compact discrete sequences, greatly compressing the informational complexity of the data. We then introduced efficient hierarchical Transformers with sequence-masking, in order to model a factorization of the joint probability of the top and bottom \vqvaetwo{} codemaps. These models are trained to perform autoregressive prediction, but incorporate information both form the past and the future of the sequences, so that they can be used to perform inpainting on the codemaps at inference time. We introduced a web interface that allows to use the trained models to generate and edit sounds, which we distribute in an open-source fashion, for researchers and musicians.

Future work includes speeding up and scaling the model for better reconstruction and faster generation, via, for instance, recent advancements to Transformer-like architectures~\cite{katharopoulos20:transformers_are_rnns_fast_autoregressive_transformers_with_linear_attention}, along with a more in-depth, quantitative assessment of the performances of the proposed framework and architectures.

\bibliographystyle{apacite}
\bibliography{bibliography.bib}

\newpage

\appendix
\markboth{Appendix}{Appendix}

\section{Custom representation and initialization}

We detail here the various changes we have made to the default settings for both the \gansynth{} representation and the \vqvae{}. These empirically provided better results during our experiments, although a proper ablation study remains to be done. 

\subsection{Phase thresholding for robust \vqvaetwo{} Training\label{subsubsec:phase_thresholding}}

In~\cite{engel19:iclr:gansynth}, the authors note that the phase of spectrograms becomes practically random at low sound amplitudes. This is highly undesirable in our context since the \vqvae{} is expected to learn a small number of codewords to represent a very diverse amount of timbral templates. The perceptually irrelevant noise in the IF maps could therefore distract it from learning appropriate representations. We consequently set an amplitude threshold and clamp the amplitude of the spectrograms to this threshold, simultaneously setting the phase to \( 0 \) in the Mel-IF representation wherever the amplitude falls beyond the threshold. The threshold's value is set according to the training dataset's dynamics. We observed that the resulting models learn a single code for representing zero amplitude zones in the spectrograms, resulting in homogeneous codemaps in case of silence. 

Since the aim of this operation is to help the \vqvae{} not be distracted by the irregular phase values during training, we add the same masking transformation at the output of the decoder, so that the model always generates valid, phase masked spectrograms so that no gradients are propagated regarding these areas and no time is spent during training to learn to reconstruct these zones.

\subsection{Increasing the Mel-scale's non-linearity}

We finally slightly adapt the Mel-IF representation used by \gansynth{} to make it more compatible with our inpainting-based approach. Namely, we lower the Mel scale's \emph{break frequency}, which defines the transition from a linear to logarithmic scale, from \( 700Hz \) to \( 240Hz \) in order to obtain more resolution in the lower range, resulting in the following Hertz-to-Mel conversion formula: \( f_{Mel} := Q * \log{f_{hZ} / 700}\), with \( Q \) a normalization constant. Without this modification, we observed that most of the frequency scale, was devoted to frequencies above \( 700Hz \). Since we perform very coarse downsampling in the \vqvae{}'s encoder, the resulting codemaps are typically comprised of \( 16 \) to \( 64 \) linearly spaced frequency bands over the spectrogram. Such a representation over a mostly linear frequency-axis would therefore lead to most of the lower end of the spectrum up to the mid frequencies being inconveniently condensed on a very limited number of bands.

\subsection{Codebook Initialization}

\vqvae{}s are experimentally unstable in training and codewords tend to fluctuate a lot due to the highly-non-linear discrete assignment the training loss relies on. This is especially true at the beginning of training, where codewords can experimentally be seen to successively explode before decreasing again in amplitude.
In order to mitigate this strong irregularity at the beginning of training, we initialize the \vqvae{}'s codewords based on the initial variance of outputs of the encoder. We empirically set this value to 0.001 based on several random initializations of the models.

\section{Spectrogram Transformers}

For these Transformers, we adopt the conditioning and masking schemes described in~\cite{hadjeres20:vqcpc} for efficient \emph{template-based} generation of symbolic music using sequence-masked, hierachical Transformers. Our setting is indeed coherent with theirs: the top codemap can be seen as a template defining the global structure of the sound, and the underlying bottom map is generated conditioned on this template. An encoder Transformer is then used to sum up the template and a decoder Transformer generates the bottom codemap whilst attending to information from the top codemap through cross-attention. Our approach nonetheless extends from theirs since we also learn to generate the templates (the top codemaps) in an autoregressive fashion whilst the authors of~\cite{hadjeres20:vqcpc} extract templates from existing music to subsequently generate variations thereof.

\subsection{Self-Attentive Top Transformer}

The top Transformer learns to model the distribution \( p(c_{T}) \) through the factorization \( p(c^{T}_i|c^{T}_{<i}, m \bigodot{} c^{T}_{\geq{}i}) \), where \( m \) is the \emph{inpainting mask}, a Boolean mask that allows the model to support interactive regeneration by inpainting. Indeed, in order to perform inpainting on arbitrary zones in the top codemap, the Transformer must be able to account for both past and future \emph{constraints}, that is, fixed tokens outside of the inpainted zone that will not be resampled during the inpainting step. To this end, we use a self-conditioned encoder-decoder architecture for the top Transformer. Here, the source -- the input to the encoder -- is the top codemap, masked by an anti-causal mask augmented with the inpainting mask \(m\): tokens that are to be inpainted in the future are not to be accounted for in the autoregressive sampling and are therefore masked. During training, we randomly sample encoder inpainting masks from a Bernoulli distribution so as to simulate the partial-information behavior that the model is faced with at inference time. The target -- the input to the decoder -- is the causally-masked top codemap.  For a given token, the encoder sums up information about the past for consistent autoregressive generation and the decoder aggregates future constraints to ensure that the generated tokens be compatible with these future constraints. This anti-causal encoder, causal decoder architecture is in line with the Anticipation-RNN architecture for inpainting described in~\cite{hadjeres18:anticipation_rnn}.

For sequence modeling operations, the top codemaps are readily flattened in column-major order, in the direction of ascending frequencies, that is, the top Transformers are trained to generate sounds time-frame by time-frame, starting from the lowest frequencies and going up, which we found to be consistent with the harmonic structure of musical sounds.

\subsection{Top-Conditioned Bottom Transformer}

The bottom Transformer models \( p(c_{B}|c_{T}) \) via the factorization \( p(c^{B}_i|c^{B}_{<i}, c^{T}_{k}) \), with \( k \) the index of the patch in the top code-map aligned with token \( i \) in the bottom codemap. We use a similar causal decoder, anti-causal encoder architecture for this top-conditioned, bottom Transformer, but we further take into account the strong structural conditioning induced by the hierarchical, patch-wise structure of the \vqvae{}. We therefore restrict the cross-attention mechanism to only allow tokens of the bottom codemap to attend to information from their "parent" patch in the conditioning top sequence, that is, we use a diagonal cross-attention mask. This allows to reduce the computational cost of the cross-attention from quadratic to just linear.

For the bottom codemaps, we must take into account the top-codemap conditioning during the linearization, so as to properly align the top and bottom sequences. Since the bottom codemap is an upsampled version of the top codemap, each token of the top codemap essentially conditions a \emph{patch} of the bottom codemap. The dimension of patches is \( D_F \times D_T \), with \( D_F \)  and \( D_T \) the downsampling ratios respectively in frequency time and time from the bottom codemap to the top codemap. We linearize the bottom map according to the order of patches defined by the linearization of the top codemap. We then linearize each \( D_F \times D_T \)-sized patch of the bottom codemap in the order we used for the top codemap, so that both Transformers locally operate identically. This results in a zig-zag linearization of the bottom, both informed by the harmonic structure of sound and coherent with the hierarchical top-to-bottom conditioning scheme

Transformers trained to perform next-token prediction commonly use an added \emph{start symbol} to prime the model in a normalized fashion. In order to properly align the top and bottom sequences in terms of patches, we expand the number of start symbols in the bottom sequences to \( D_F \times D_T \). The top start symbol can therefore be interpreted as a "first patch" token with corresponding, aligned start tokens in the bottom codemap.

\subsection{Factorized Positional Embeddings}

We additionally use factorized positional encodings to introduce time-frequency information in the linearized models. These small embeddings are appended to the individual tokens in the linearized sequences and provide the Transformers with location along these sequences. For the top codemaps, the positional embedding of a token is the Cartesian product of its time index and the index of its frequency band, providing an inductive bias for both global temporal structure and local harmonic structure. For bottom codemaps, we identify tokens by combining a frequency embedding for the frequency band of it's parent patch's and a patch-wide positional embedding, which only distinguishes position within a given patch. The bottom decoders are therefore effectively unaware of the global temporal structure of the codemaps, which we provide only to the top encoder. This is aligned with our goal of modeling large-scale structure via the top codemap and local structure via the bottom map. Finally, we augment the positional embeddings for both the top and bottom sequences with one-hot embeddings for the pitch and instrument class, to allow the models to take this information into account. Other kinds of global conditioning can be added here in the same fashion.

\section{Implementation details}

\subsection{VQ-VAE-2}

For the hierarchical \vqvaetwo{}, we adapt the existing PyTorch implementation by user \texttt{rosinality}~\cite{rosinality:github:vq-vae-2-pytorch}, and mostly keep default parameters for the network's structure. We only adapted the existing code to support coarser downsampling and upsampling ratios. For both the top and bottom quantized layers, we use codebooks with 512 individual codewords of dimension 64 each. The downscaling scheme is as follows: the input, 4 seconds-long spectrogram, of dimension \( 1024 \times 128 \), is first downsampled via strided convolutions by a factor 16, yielding a bottom map of dimension \( 32 \times 8 \). The resulting codemaps are then further downsampled by a factor \( 2 \) resulting in a top codemap of dimension \( 16 \times 4 \).

We train the \vqvaetwo{} with \textsc{EMA} using a decay of \( 0.99 \) and we set the latent loss ratio \( \beta \) to \( 0.25 \). We use the \textsc{ADAM} optimizer with a learning rate of \( 3e-4 \). Reconstruction quality from the autoencoder is evaluated by a simple MSE loss on the spectrograms.

We trained the model for 3 days using a batch size of 256 spread over 2 NVIDIA GeForce GTX 1080 Ti GPUs. The resulting model obtains average perplexities of 270 for the bottom and top code assignations, which shows good usage of the codebooks in average (both comprised of 512 individual codewords). We generate the Mel-IF spectrograms on-the-fly on GPU. We found this to be more convenient and portable than pre-rendering the whole NSynth dataset, whilst inducing no significant overhead.

\subsection{Transformers}

We adapt the Transformer implementation proposed for the \textsc{VQ-CPC} model for template-based music generation \cite{hadjeres20:vqcpc}, using the code provided along with the paper. We use positional embeddings of dimension 32, factorized into two components of dimension 16 each. We use two separate embeddings of dimension 32 each for the pitch and instrument type modalities, which are appended to the positional embeddings of each token. Finally, we embed the \vqvae{}'s token into a space of dimension \( 416 \) for an effective model size of \( 512 \).
The top and bottom transformers both use \( 6 \)-layer encoders, \( 6 \)-layer decoders all with \( 8 \) attention heads.

We train the Transformers to optimize the Cross Entropy using \textsc{RAdam}~\cite{liu2019:iclr:radam} with an initial learning rate of \( 10^{-4} \), since this optimizer was shown to work well for training Transformers. We use gradient clipping with clipping value \( 5 \). We furthermore perform label smoothing~\cite{muller19:neurips:when_does_label_smoothing_help}, replacing the discrete targets of the next-token prediction task with partially noisy ones. To do so, we distribute a mass of \( 0.05 \) from the probability of the actual ground-truth token uniformly across all the other possible tokens. This was shown to help train multi-class models for autoregressive generative tasks. During training of the self-conditioned Top transformer, in order to limit the information available to the decoder, whilst letting it learn to account for future constraints, we sample encoder masks for each training sequence independently from a Bernoulli distribution. The probability of this Bernoulli is sampled for each batch between \( 0.8 \) and \( 1. \), effectively varying the amount of conditioning information the decoder is given access to.
At inference time, we sample from the autoregressive models using top-\(p\) sampling with \( p = 0.8 \), which was shown to improve the quality of autoregressive sampling with Transformer models~\cite{holtzman20:iclr:the_curious_case_of_neural_text_degeneration_aka_nucleus_sampling_top_p_sampling}.

Both models can be trained independently, since they only depend on the codemaps produced by the \vqvaetwo{}. We pre-compute these compact representations and store them in an efficient-access database, to alleviate the cost of loading each 4 seconds-long sounds in succession.

We train the top Transformer for 2.5 days with a batch size of 280 spread accross 4 GeForce GTX 1080 Ti GPUs.
We train the bottom Transformer for 2.5 days with a batch size of 80 over 4 GeForce GTX 1080 Ti GPUs.

\end{document}